\documentclass[twocolumn,showpacs,preprintnumbers,amsmath,amssymb,prb]{revtex4}
\usepackage[dvips]{color,graphics,epsfig,rotating}
\usepackage{graphicx}
\usepackage{dcolumn}
\usepackage{bm}

\begin{document}

\title{Electronic structure and Magnetism in BaMn$_2$As$_2$ and BaMn$_2$Sb$_2$}

\author{Jiming An}
\affiliation{Materials Science and Technology Division,
Oak Ridge National Laboratory, Oak Ridge, Tennessee 37831-6114} 
\affiliation{Wuhan University of Technology, Wuhan, China}

\author{A.S. Sefat}
\affiliation{Materials Science and Technology Division,
Oak Ridge National Laboratory, Oak Ridge, Tennessee 37831-6114} 

\author{D.J. Singh}
\affiliation{Materials Science and Technology Division,
Oak Ridge National Laboratory, Oak Ridge, Tennessee 37831-6114} 

\author{Mao-Hua Du}
\affiliation{Materials Science and Technology Division,
Oak Ridge National Laboratory, Oak Ridge, Tennessee 37831-6114} 

\date{\today} 

\begin{abstract}
We study the
properties of ThCr$_2$Si$_2$ structure BaMn$_2$As$_2$ and
BaMn$_2$Sb$_2$ using density functional calculations
of the electronic and magnetic as well 
experimental measurements on single crystal samples of BaMn$_2$As$_2$.
These materials are local moment magnets with moderate band gap
antiferromagnetic semiconducting ground states.
The electronic structures show substantial
Mn - pnictogen hybridization, which stabilizes an intermediate spin
configuration for the nominally $d^5$ Mn.
The results are discussed in the context of possible thermoelectric
applications and the relationship with the corresponding iron / cobalt / nickel
compounds Ba(Fe,Co,Ni)$_2$As$_2$.
\end{abstract}


\maketitle

\section{introduction}

Requirements for a thermoelectric material to be used in waste
heat recovery include a high figure of merit,
$ZT=\sigma S^2 T/\kappa$, preferably unity or higher,
and materials cost compatible with the application.
Here $S$ is the Seebeck coefficient
(thermopower), $\sigma$ is the electrical conductivity,
and $\kappa$ is the thermal conductivity.
High $ZT$ cannot be obtained without high thermopower.
\cite{s-note}
Current materials for intermediate temperatures typical of 
requirements for automotive exhaust heat conversion
contain substantial amounts of Te, as in e.g. PbTe.
However, Te is a rare element whose supply is mainly as a byproduct
of Cu production and
whose cost is expected to sensitive to usage. In particular,
this may lead to difficulties in
applications that require large volumes of Te based thermoelectric
material.
Therefore it is of interest to identify alternative materials based
on common inexpensive elements. Among metals and metalloids
these include alkaline earths (Mg,Ca,Sr,Ba), mid-3 $d$ transition elements,
especially Mn and Fe and pnictogens (As,Sb).
Wang and co-workers \cite{wang}
recently reported measurements of the transport properties of single
crystals of BaMn$_2$Sb$_2$ grown from Sn flux. They report a
room temperature thermopower of $\sim$ 225 $\mu$V/K,
but accompanied by a low electrical conductivity leading to low $ZT$. The
conductivity increases with temperature, but this is accompanied by
a decreasing thermopower.
This suggests intrinsic semiconducting behavior, which is consistent with
activated fits of the conductivity. Interestingly, a transition was found from
a low activation energy of $\sim$ 0.1 eV ($T<$470 K)
to a higher activation energy
region ($\sim$ 1.1 eV) at $T>$650 K, possibly reflecting a cross-over
from an extrinsic to an intrinsic regime.
However, doping studies have not yet been reported.

Recent discoveries of superconductivity \cite{kamihara,rotter}
in ThCr$_2$Si$_2$ structure
pnictides has led to renewed interest in this large class of compounds.
\cite{pearson,just}
Among the compounds $AM_2Pn_2$, $A$=Ca,Sr,Ba,Eu, $Pn$=As,
high temperature superconductivity is found for $M$=Fe, and an
apparently different superconductivity for $M$=Ni.
\cite{ronning-bna,kurita-bna,subedi-bna}
The $M$=Co compound, BaCo$_2$As$_2$ is near ferromagnetism,
\cite{sefat-bca}
while CaCo$_2$P$_2$
is reported as having ferromagnetically ordered Co planes, which are
stacked antiferromagnetically. \cite{reehuis}
Interestingly, unconventional superconductivity is found in
Ba(Fe$_{1-x}$Co$_x$)$_2$As$_2$ for a substantial range of $x$ around 0.1.
\cite{sefat}
Turning to the electronic structures, the compounds with $M$=Fe,Co and Ni
show roughly similar shapes of their densities of states, with the main
difference being the position of the Fermi energy, due to the different
electron counts. These electronic structures show moderate hybridization
between the transition metal $d$ bands and the As $p$ bands, with the
main As $p$ bands $\sim$ 2-3 eV below the Fermi energy, $E_F$ and
approximately $\sim$ 10-20\% As characater in the $d$ bands. Thus the
As occurs nominally as As$^{3-}$. Interestingly, even though ThCr$_2$Si$_2$
is an very common structure type, the $Pn$=Sb compounds, namely BaFe$_2$Sb$_2$,
BaCo$_2$Sb$_2$ and BaNi$_2$Sb$_2$ are not reported.
On the other hand, the Mn compounds, BaMn$_2$P$_2$, BaMn$_2$As$_2$
and BaMn$_2$Sb$_2$ are known phases as are the corresponding Ca, Sr and Eu
materials, though with structural distortions in some cases.
\cite{brechtel-bms,payne,bobev,brock-bmp,xia-bms}

From a practical point of view,
there are several reports of remarkably high thermopowers
accompanied by metallic conduction in the Fe based superconducting phases.
\cite{sefat-s,pinsard,mcguire,li,li-ba}
These materials have relatively narrow manifolds of Fe $d$ bands
near the Fermi energy. These result from the combination
of moderate Fe-As hybridization and expanded Fe-Fe distance as compared with
bulk Fe.
In any case,
the values of $S$ in the Fe compounds
are too low by a factor of $\sim$ two compared with
practical thermoelectrics, and are high at low temperatures, inapplicable
to waste heat recovery.
As mentioned, BaMn$_2$Sb$_2$ is reported to be a semiconductor.
\cite{wang,xia-bms}
We note that BaMn$_2$Sb$_2$,
BaMn$_2$As$_2$ and BaFe$_2$As$_2$ are composed 
of common, inexpensive elements, which as mentioned
is an important consideration
for some thermoelectric applications.
Key questions concern the relationship between BaMn$_2$As$_2$ and
BaMn$_2$Sb$_2$, and that between the Mn and Fe compounds, as well
as the related issue of whether the conductivity in the Mn compounds
can be improved by doping, while at the same time maintaining
high thermopowers.

\section{theoretical methods}

The purpose of this paper is to further elucidate the electronic properties
of these Mn compounds in relation to the Fe phases, in particular
by comparing BaMn$_2$As$_2$ with BaFe$_2$As$_2$, and with BaMn$_2$Sb$_2$,
which was recently studied by Xia and co-workers. \cite{xia-bms}
We find that
the Mn compounds are rather different from the Fe, Co and Ni compounds
in that there is much stronger hybridization with the pnictogen $p$ states.
This leads to an electronic structure that cannot be regarded as
derived from that of the Fe compound with a different electron count.
Furthermore, in contrast to the metallic character and itinerant
magnetism in the Fe phases, we find local moment, semiconducting behavior
in both BaMn$_2$Sb$_2$ and BaMn$_2$As$_2$.

Electronic structure calculations were performed within density functional
theory.
We did calculations using both the local density approximation (LDA)
and the generalized gradient approximation (GGA)
of Perdew and co-workers, \cite{pbe}
and also using both the
general potential linearized augmented planewave (LAPW) method
\cite{singh-book}
similar to our earlier calculations for the Fe-based materials,
\cite{singh-lfao,singh-bfa}
and an ultrasoft pseudopotential method implemented in the
Quantum Espresso package. \cite{pwscf}
For the LAPW calculations, we used two different codes, an in-house code
and the Wien2k code, \cite{wien2k}
and compared results, finding no significant differences.
We used
LAPW sphere radii of 2.1 Bohr for the elements other than Ba
in BaMn$_2$As$_2$ and 2.2 Bohr for all elements in BaMn$_2$Sb$_2$.
Two different sphere radii were used for Ba in BaMn$_2$As$_2$:
2.35 Bohr with Wien2k and 2.2 Bohr in the other LAPW calculations.
In all cases, converged basis sets, corresponding to $R_{min}k_{max}$=9
with additional local orbitals \cite{singh-lo}
were used, where $R_{min}$ is the 
minimum LAPW sphere radius and $k_{max}$ is the planewave cutoff.
Convergence of the Brilloiun zone sampling was checked by directly
calculating results using different grids.
We used the literature values of the lattice parameters,
$a$=4.418\AA, $c$=14.200\AA, for BaMn$_2$Sb$_2$ and
$a$=4.15\AA, $c$=13.47\AA, for BaMn$_2$As$_2$ with internal
parameters, $z_{As}$ and $z_{Sb}$ determined by LDA total energy minimization
in the low energy antiferromagnetic state. The resulting values were
$z_{Sb}$=0.3594 for BaMn$_2$Sb$_2$ and $z_{As}$=0.3524 for BaMn$_2$As$_2$.
The value obtained for BaMn$_2$Sb$_2$ is in reasonable
accord with the experimental value \cite{xia-bms} of 0.3663.

Transport calculations were performed within the constant scattering
time approximation using the BoltzTraP code. \cite{boltztrap}
For consistency we focus here on results obtained within the LDA
with the LAPW method, but we note that the conclusions are robust
with very similar results and conclusions
emerging from the GGA planewave calculations.
We also note that our electronic structure for BaMn$_2$Sb$_2$ is
quite similar to that reported previously by Xia and co-workers.\cite{xia-bms}

\section{synthesis}

Experimental characterization of BaMn$_2$As$_2$ was performed using
single crystals. These were grown out of MnAs flux. The typical
crystal sizes were $\sim$ 5x3x0.2 mm$^3$. High purity elements
($>$ 99.9\%) were used in the preparation of the samples. The
source was Alfa Aesar.
First, MnAs binary was prepared by placing mixtures of As, Mn pieces
in a silica tube.  These were reacted slowly by heating to 300 $^\circ$C
(25 $^\circ$C/hour, dwell 10 hours), and 
to 600 $^\circ$C (15 $^\circ$C/hour, dwell 75 hours).
Then a ratio of Ba:MnAs = 1:5 was heated in an alumina crucible for 15 hours
at 1230 $^\circ$ C under partial atm argon. This reaction was cooled at the
rate of 2 $^\circ$/hour, followed by decanting of the MnAs flux at
1120 $^\circ$C.

\section{crystal structure, transport and specific heat}

The crystals were malleable but well-formed plates with the [001]
direction perpendicular to the large faces. The structural identification was
made via powder x-ray diffraction using a Scintag XDS 2000
$\Theta$-$\Theta$ diffractometer (Cu $K_\alpha$ radiation).
The cell parameters were refined using least squres fitting of the
measured peak positions in the range 2$\Theta = 10 - 90 ^\circ$
using the Jade 6.1 MDI package. The resulting lattice parameters
of BaMn$_2$As$_2$ were $a$=4.1633(9) \AA, $c$=13.448(4) \AA, in
the ThCr$_2$Si$_2$ structure
($I4/mmm$, $Z$=2). The cell volume for BaMn$_2$As$_2$
is then 233.087(4) \AA$^3$, which is in accord with
prior reports and much larger than BaFe$_2$As$_2$ 
(204.567(2) \AA$^3$) (Ref. \onlinecite{sefat})
or BaCo$_2$As$_2$
(197.784(1) \AA$^3$) (Ref. \onlinecite{sefat-bca}).

\begin{figure}[tbp]
\vspace{0.2cm}
\epsfig{file=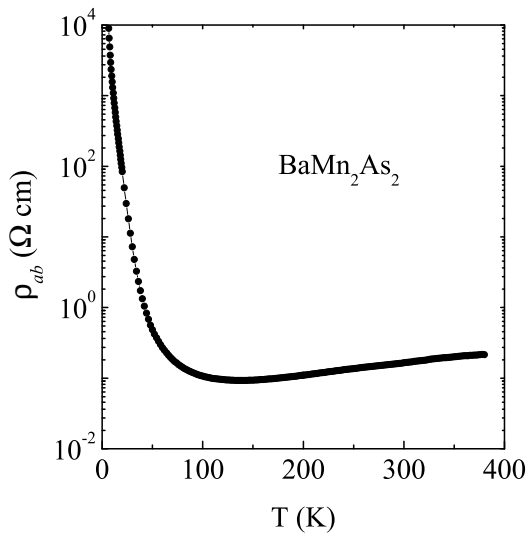,width=0.90\linewidth,angle=0,clip=}
\caption{Temperature dependence of the in-plane resistivity of
BaMn$_2$As$_2$ betrween 1.8 K and 380 K.}
\label{rho}
\end{figure}

Temperature dependent electrical resistivity measurements were
performed on a Quantum Design Physical Property Measurement System (PPMS),
measured in the $ab$-plane above 5 K. For BaMn$_2$As$_2$, the
resistivity, $\rho$ decreases with increasing temperature up to 
$\sim$ 150 K, above which it gives a decreasing, metallic-like
dependence (Fig. \ref{rho}).
The resistivity at room temperature is 165 m$\Omega$ cm.

\begin{figure}[tbp]
\vspace{0.2cm}
\epsfig{file=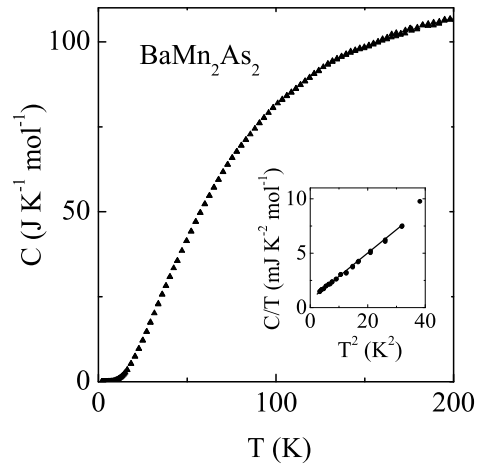,width=0.80\linewidth,angle=0,clip=}
\caption{Temperature dependence of the specific heat for
BaMn$_2$As$_2$. The inset shows $C/T$ vs. $T^2$ and a linear
fit between 1.8 K and 3.5 K}
\label{cp}
\end{figure}

Specific heat data, $C_p(T)$, were also obtained using the PPMS via
the relaxation method.
Fig. \ref{cp} gives the temperature dependence of the specific heat.
There are no features suggesting a phase transition in the specific
heat up to 200 K. Below $\sim$ 6 K, $C/T$ has a linear $T^2$ dependence
(inset of Fig. \ref{cp}). The fitted Sommerfeld coefficient, 
$\gamma$=0.79(2) mJ/(K$^2$ mol), or $\sim$ 0.4 mJ/(K$^2$ mol Mn) is
near zero.
For comparison the
value fo BaFe$_2$As$_2$, which has a spin density wave instability gapping
most of the Fermi surface is 3.0 mJ/(K$^2$ mol Fe), \cite{sefat}
and the
value for BaCo$_2$As$_2$ is 20.8 mJ/(K$^2$ mol Co). \cite{sefat-bca}.
In fact, it is likely that the small observed $\gamma$ arises from
a small amount of MnAs flux included in the sample and is therefore
not intrinsic to BaMn$_2$As$_2$. This is based on
squid measurements, where we find a small signal at 310 K, corresponding
to the ordering temperature of MnAs.
The value for the Debye temperature is
$\Theta_D \sim$ 280 K, estimated above 150 K, using the calculated values
of the $T/\Theta_D$ dependence of the Debye specific heat model.
Thus our transport and specific heat measurements are consistent with
a small band gap semiconductor. They are not consistent with either a
large gap insulator (e.g. a Mott insulator)
or a metal with significant carrier density.

\section{electronic structure and magnetism}

We begin the discussion of the density functional investigation
with a summary of our calculated results for BaMn$_2$Sb$_2$,
which as mentioned are similar to those obtained previously by Xia and
co-workers, \cite{xia-bms}
although in general the LAPW method is more accurate than the 
linear muffin tin orbital method employed in that work.
Calculations were performed for a checkerboard antiferromagnetic
ordering (AF-C), which is the lowest energy state, as well as a
ferromagnetic ordering (F), and zone boundary antiferromagnetic
orders alternating chains of like spin Mn (as in the spin
density wave state of BaFe$_2$As$_2$),
and a non-spin-polarized (NSP) case.
The two cells with chain ordering are denoted AF-S1, which
consists of chains stacked antiferromagnetically along the $c$-axis
direction, and AF-S2, with ferromagnetic stacking.
The energetics are summarized
along with those of BaMn$_2$As$_2$ in Table \ref{tab-mag}.
We find a strong tendency towards moment formation, which results
from the strong Hund's rule coupling on the nominal Mn$^{2+}$ ions
in the compound, with stable moments independent of the magnetic order.
Integrating within the Mn LAPW sphere (radius 2.2 Bohr) we obtain moments
of 3.47 $\mu_B$ for the ferromagnetic (F) ordering and 3.55 $\mu_B$ for the
G-type checkerboard AF-C ordering, which is the lowest energy state
(n.b. G-type ordering is the experimental ground state at least
for the phosphide). \cite{brock-bmp}
Including contributions from all atoms in the unit cell, the
F ordered spin magnetization is 3.76 $\mu_B$ on a per Mn basis,
reflecting a small parallel contribution from the Sb.
Significantly, this is considerably reduced from the nominal
5 $\mu_M$ expected for high spin Mn$^{2+}$. This reduction
reflects strong hybridization between Mn $d$ and Sb $p$ states
in this compound. In general, very strong hybridization is needed
to effectively compete against the very strong Hund's interaction
in the half-filled $d$ shell of Mn$^{2+}$ though we note that
strong hybridization was previously identified in
the phosphides by Hoffmann and Zheng and by Gustenau and co-workers.
\cite{hoffmann,zheng,gusteneau}
Also, associated with the strong hybridization we find a very large
energy difference between the ferromagnetic and antiferromagnetic
orderings. This amounts to more than 0.29 eV/Mn in the LDA and implies
a very high magnetic ordering temperature well above room temperature.
This would be similar to the phosphide, which is reported to have
an ordering temperature above 750K. \cite{brock-bmp}
For comparison the corresponding energy for bcc Fe (Curie temperature,
$T_C$=1043K)
is 0.4 eV, with opposite (ferromagnetic) sign. \cite{singh-fe}

\begin{table}
\caption{Magnetic energies in eV of BaMn$_2$Sb$_2$ and BaMn$_2$As$_2$
as obtained within the LDA on a per formula
unit (two Mn) basis. The non-spin-polarized case, denoted
NSP, is set as zero. F denotes ferromagnetic order, while AF-C is the
checkerboard G-type antiferromagnetic ordering, and AF-S1 and
AF-S2 are stripe-like orders (see text).}
\label{tab-mag}
\vspace{0.3cm}
\begin{tabular}{|l|c|c|}
\colrule
    &   ~BaMn$_2$Sb$_2$~   & ~BaMn$_2$As$_2$~ \\
\colrule
~NSP~  &    0  & 0 \\
~F     & -1.55   & -0.66 \\
~AF-S1 & -1.94   & -1.03 \\
~AF-S2 & -1.91   & -1.01 \\
~AF-C  & -2.14   & -1.32 \\
\colrule
\end{tabular}
\end{table}

\begin{figure}[tbp]
\vspace{0.3cm}
\epsfig{file=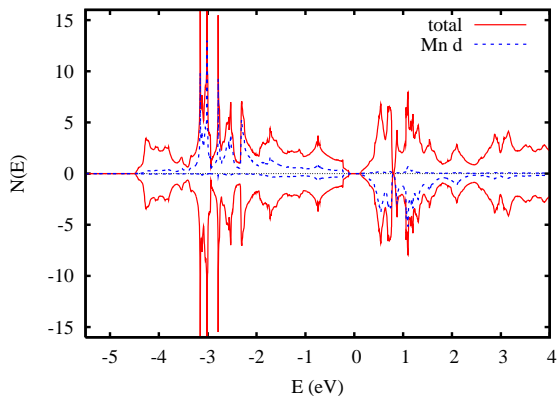,height=0.90\linewidth,angle=270,clip=}
\caption{(color online) Electronic DOS and projections
onto the LAPW spheres for antiferromagnetic BaMn$_2$Sb$_2$. The plot
shows the total DOS above and below the axis, the Mn majority spin
projection above and the Mn minority spin projection below.
The remaining (non-Mn) contribution to the valence bands is mainly from
Sb $p$ states.}
\label{dos-bmaf}
\end{figure}

In fact, strong, spin dependent hybridization is seen in the electronic
structure, represented by the electronic density of states and projections
(Fig. \ref{dos-bmaf}). As may be noted, the antiferromagnetic structure
is found to be a small band gap semiconductor, $E_g$=0.21 eV in the LDA.
The majority spin Mn $d$ states overlap in energy with the Sb $p$ states
and are strongly mixed with them. The minority spin Mn $d$ states
are above the main Sb $p$ DOS, and are therefore less strongly mixed
with the Sb states. Nonetheless, there is sufficient minority spin Mn
$d$ - Sb $p$ hybridization to lead to noticeable Mn $d$ character in the
occupied valence band states. This is the main source of the moment
reduction, corresponding to a Mn $d$ electron count closer to Mn$^{1+}$
than Mn$^{2+}$. Finally we note that such strong spin dependent
hybridization while leading to strong superexchange interactions
is highly unfavorable for carrier mobility in an antiferromagnet,
\cite{pwa}
since at low $T$ hopping between nearest neighbor
sites with opposite spin will be suppressed and at higher $T$ spin disorder
will lead to strong scattering.

We now discuss our experimental and calculated results for BaMn$_2$As$_2$.
The calculated magnetic energies (Table \ref{tab-mag}) show strong local
moment magnetism with a large exchange, similar to BaMn$_2$Sb$_2$, with the
exception that the energies of the magnetically ordered states are
not as low compared with the non-spin-polarized case, as those in the
antimonide. This reflects lower moments due to even stronger hybridization.
The LDA total spin magnetization for the F ordering is 2.9 $\mu_B$/Mn,
i.e. $\sim$ 0.9 $\mu_B$ lower than in the antimonide. The moment inside
the 2.1 Bohr LAPW sphere is 3.20 $\mu_B$ for the AF ordering and 2.74
$\mu_B$ for the F ordering. Nonetheless, the ordering energy remains
very high, again reflecting strong spin-dependent hybridization.
The energy difference between the AF and F orderings is 0.33 eV / Mn
in the LDA. The corresponding GGA value of 0.38 eV is similar, the main
difference between the LDA and GGA results being that the moment
formation is stronger in the GGA, leading to moments in the LAPW 
spheres of 3.07 $\mu_B$ for F ordering and 3.35 $\mu_B$ for AF ordering.
Turning to the AF-S1 and AF-S2 orders, the energies are slightly lower
than the average of the F and AF-C states. A simple nearest and next
nearest neighbor superexchange picture is probably an oversimplification
for these narrow gap semiconductors. However, the results indicate that
the nearest neighbor interaction is dominant over further interactions
and that the $c$-axis interaction is considerably weaker than the in-plane
interaction and that it is antiferromagnetic.

\begin{figure}[tbp]
\vspace{0.3cm}
\epsfig{file=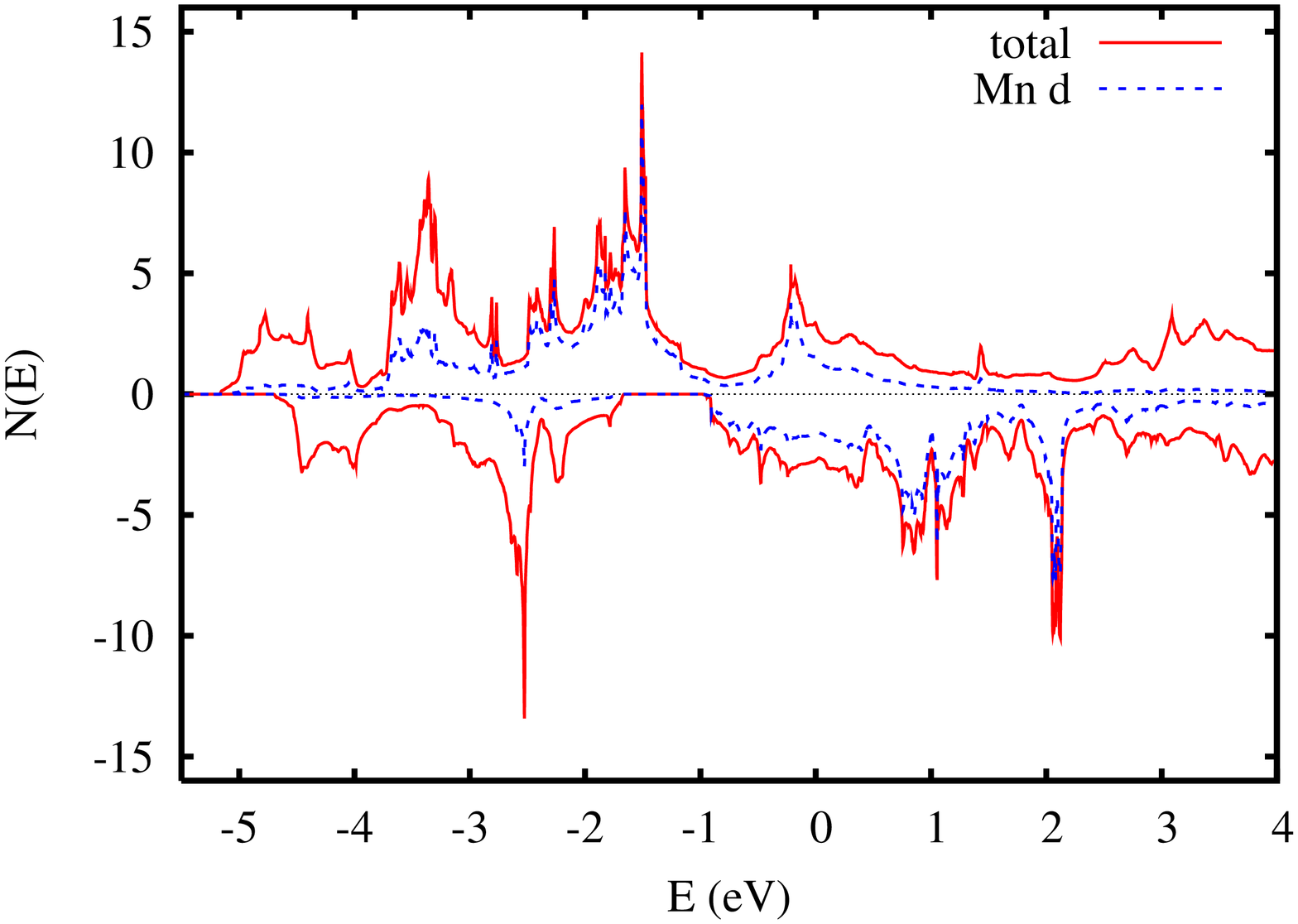,height=0.90\linewidth,angle=270,clip=}
\epsfig{file=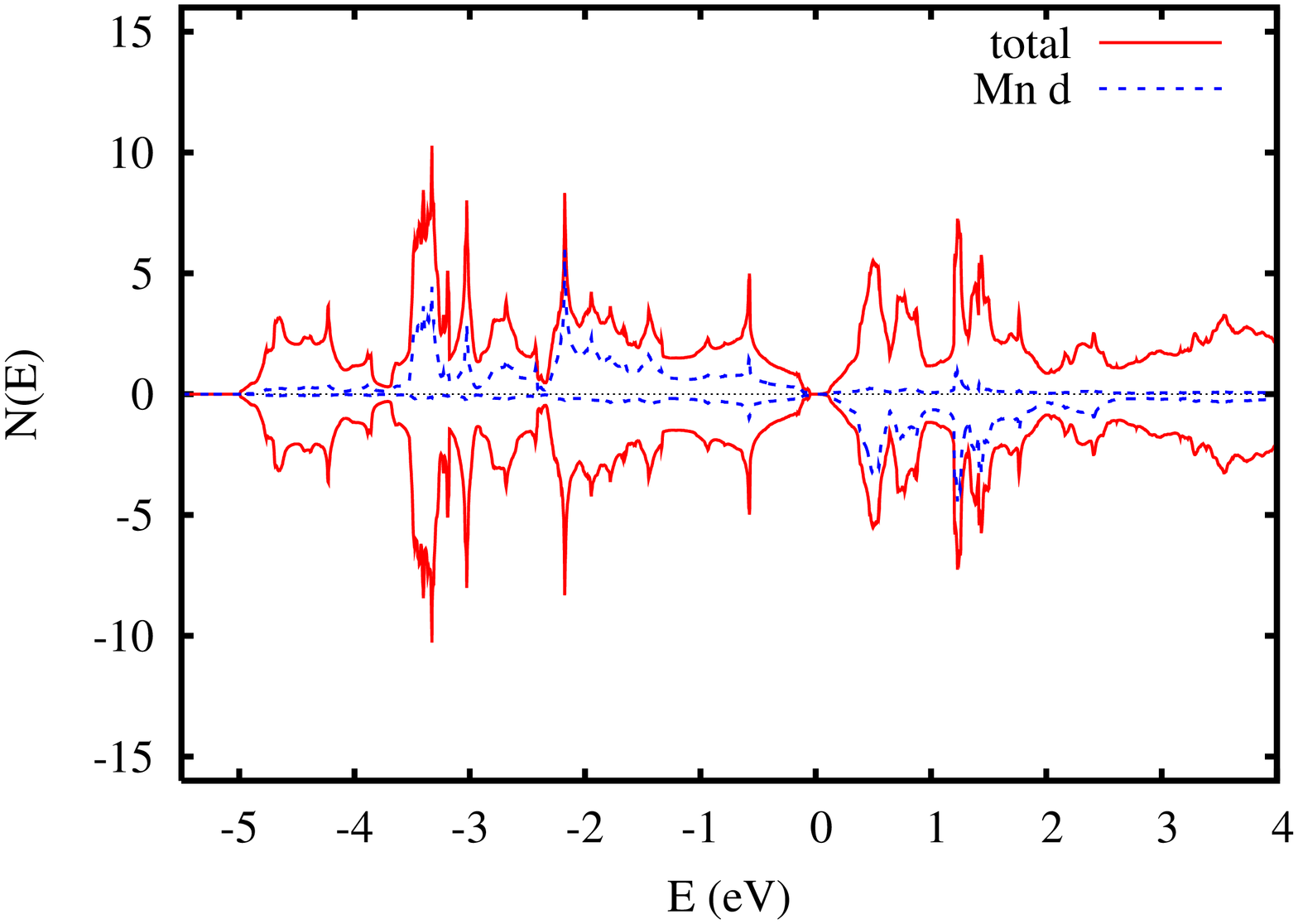,height=0.90\linewidth,angle=270,clip=}
\caption{(color online) Electronic DOS and projections
onto the LAPW spheres for ferromagnetic ordered (top) and
antiferromagnetic BaMn$_2$As$_2$ (bottom). The plots show majority spin
above the axis and minority spin below.
The remaining (non-Mn) contribution to the valence bands is mainly from
As $p$ states.}
\label{dos-ba}
\end{figure}

The calculated LDA electronic density of states of BaMn$_2$As$_2$ is
shown in Fig. \ref{dos-ba} for both F and AF orderings.
The low energy AF state again has a small semiconducting gap
of 0.1 eV in the LDA and 0.2 eV in our GGA calculations.
These low values are consistent with the activation energies
from resistivity data below 470 K obtained by Wang and co-workers.
\cite{wang}
Thus we interpret their low temperature value of $\sim$ 0.1 eV
as the intrinsic bulk gap for the AF ordered state.
Within this scenario, the larger activation gap at high $T$ would be
associated with Anderson localization arising from magnetic disorder
above the AF ordering temperature.

Strong spin-dependent hybridization is seen, and in particular
there is even more minority spin Mn character mixed with the
pnictogen derived valence bands than in the antimonide.
This stronger hybridization is consistent with the general trend that
As has a greater tendency towards bond formation with transition elements
than Sb and is reflected in the higher energy difference between the
F and AF magnetic orderings in BaMn$_2$As$_2$ than in BaMn$_2$Sb$_2$.
We emphasize that this behavior is very different from that found
in BaFe$_2$As$_2$, BaCo$_2$As$_2$ and BaNi$_2$As$_2$, which
have some $d$ - $p$ hybridization, as in an oxide, but are much more
ionic.
\cite{subedi-bna,sefat-bca,singh-bfa}
The distinct Mn - As bonding in this structure also explains why
Mn is not an effective dopant for the Fe-based superconducting compounds,
while Co and Ni are.
\cite{matsuishi}

\begin{figure}[tbp]
\vspace{0.3cm}
\epsfig{file=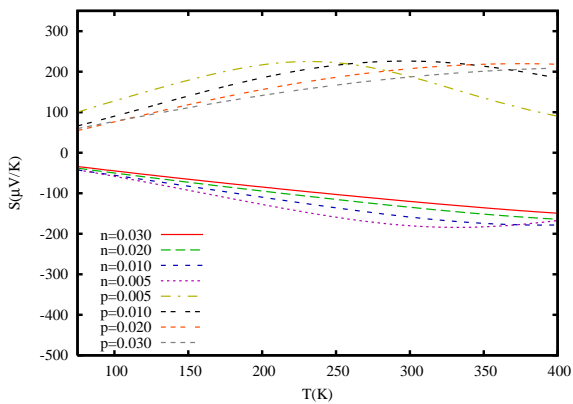,height=0.90\linewidth,angle=270,clip=}
\caption{(color online) LDA constant scattering time approximation
Seebeck coefficient in the basal plane
of AF ordered BaMn$_2$As$_2$. The doping levels
are in carriers per formula unit (2 Mn).
}
\label{sxx}
\end{figure}

\begin{figure}[tbp]
\vspace{0.3cm}
\epsfig{file=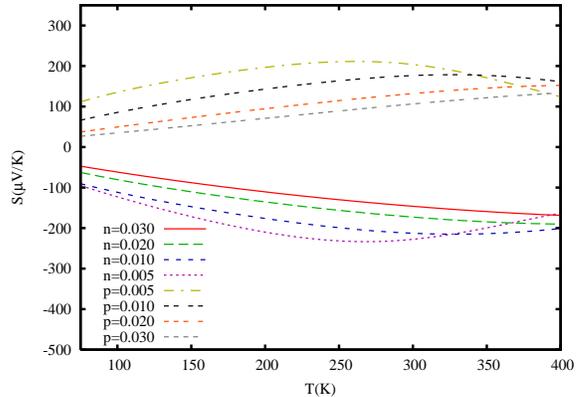,height=0.90\linewidth,angle=270,clip=}
\caption{(color online) LDA constant scattering time approximation $c$-axis
Seebeck coefficient of AF ordered BaMn$_2$As$_2$. The doping levels
are in carriers per formula unit.
}
\label{szz}
\end{figure}

\section{thermopower}

As noted, antiferromagnetism with strong spin dependent hybridization
is not favorable for high mobility conduction. Nonetheless, the
chemical flexibility of the ThCr$_2$Si$_2$ structure suggests that
high doping levels may be achievable.
In that case, if the thermopower remains high, as in the oxide thermoelectric
Na$_x$CoO$_2$, high $ZT$ may result in spite of low mobility.
\cite{terasaki}
Therefore we calculated the
Seebeck coefficients from the LDA
band structure as a function of doping level and temperature
in the constant scattering time approximation. The results are shown
in Figs. \ref{sxx} and \ref{szz}.
The anisotropies are different for the valence and conduction bands.
In the basal plane, the hole doping yields higher $S$ than electron
doping, while an opposite trend is found in the conduction bands.
This difference is due to different band anisotropies.
The constant scattering time
transport anisotropy $\sigma_{zz}/\sigma_{xx}$ estimated from
the band anisotropy is 1.9 for the valence bands ($p$-type doping)
and 0.5 for the conduction bands ($n$ type).
These modest anisotropies mean that the material is effectively a
three dimensional intermetallic compound rather than a quasi-2D system.
Turning to the values of $S$ we observe that high values compatible
with high $ZT$ are only found for modest doping levels and not for
high metallic doping levels such as those in Na$_x$CoO$_2$
(0.01 carriers per formula unit is 8.6x10$^{19}$ cm$^{-3}$;
the carrier concentration in thermoelectric Na$_x$CoO$_2$ is more than
one order of magnitude higher). \cite{gga-note}
Considering the likelihood of carrier localization in this
magnetic material,
this would initially seem not favorable for application of BaMn$_2$As$_2$
as a thermoelectric for waste heat recovery.

\section{summary and conclusions}

To summarize, we characterize BaMn$_2$As$_2$ and BaMn$_2$Sb$_2$ as
small band-gap, local moment antiferromagnetic semiconductors.
We find a very strong spin dependent hyridization. This leads to
high exchange energies, which are consistent with high magnetic
ordering temperatures. The lowest energy state found is the G-type
checkerboard antiferromagnetic state for both compounds.
We note that the bonding, electronic structure and magnetic properties
are very distinct from those of the corresponding Fe, Co and Ni
compounds, which may explain why Mn leads to carrier localization
rather than effective doping in BaFe$_2$As$_2$ and related materials.

The strong spin dependent hybridization is not favorable for
high carrier mobility. Nonetheless, the high thermopowers do suggest
that experimental
doping studies should be performed to determine the maximum $ZT$
in this material.

\acknowledgements

We are grateful for helpful discussions with B.C. Sales and D. Mandrus.
This work was supported by the Department of Energy, through
the Division of Materials Sciences and Engineering, the
Vehicle Technologies, Propulsion Materials
Program and the ORNL LDRD program.


\begin{references}

\bibitem{s-note}
This is because of the electronic contribution to thermal conductivity.
In materials where the Wiedemann-Franz relation
$\kappa_e=L\sigma T$ holds, it can be
shown that $ZT > 1$ requires $S > 160 \mu$V/K.

\bibitem{wang}
H.F. Wang, K.F. Cai, H. Li, L. Wang and C.W. Zhou,
J. Alloys Compds. doi:10.1016/j.jallcom.2008.10.080 (2008).

\bibitem{kamihara}
Y. Kamihara, T. Watanabe, M. Hirano, and H. Hosono,
J. Am. Chem. Soc. {\bf 130}, 3296 (2008).

\bibitem{rotter}
M. Rotter, M. Tegel, and D. Johrendt,
Phys. Rev. Lett. {\bf 101}, 107006 (2008).

\bibitem{pearson}
W.B. Pearson and P. Villars,
J. Less Common Met. {\bf 97}, 119 (1984);
{\em ibid.} {\bf 97}, 133 (1984).

\bibitem{just}
G. Just and P. Paufler, J. Alloys Compds. {\bf 232}, 1 (1996).

\bibitem{ronning-bna}
F. Ronnning, N. Kurita, E.D. Bauer, B.L. Scott, T. Park, T. Klimczuk,
R. Movshovich, and J.D. Thompson,
J. Phys. Condens. Matter {\bf 20}, 342203 (2008).

\bibitem{kurita-bna}
N. Kurita, F. Ronning, Y. Tokiwa, E.D. Bauer, A. Subedi, D.J. Singh,
J.D. Thompson, and R. Movshovich,
arXiv:0811.3426 (2008).

\bibitem{subedi-bna}
A. Subedi and D.J. Singh,
Phys. Rev. B {\bf 78}, 132511 (2008).

\bibitem{sefat-bca}
A.S. Sefat, D.J. Singh, R. Jin, M.A. McGuire, B.C. Sales, and D. Mandrus,
arXiv:0811.2523 (2008).

\bibitem{reehuis}
M. Reehuis, W. Jeitschko, G. Kotzba, B. Zimmer, and X. Hu,
J. Alloys Compds. {\bf 266}, 54 (1998).

\bibitem{sefat}
A.S. Sefat, R. Jin, M.A. McGuire, B.C. Sales, D.J. Singh, and D. Mandrus,
Phys. Rev. Lett. {\bf 101}, 117004 (2008).

\bibitem{brechtel-bms}
E. Brechtel, G. Cordier, and H. Schafer,
Z. Nat. {\bf 34b}, 921 (1979).

\bibitem{payne}
A.C. Payne, A.E. Sprauve, M.M. Olmstead, S.M. Kauzlarich, J.Y. Chan,
B.A. Reisner, and J.W. Lynn,
J. Solid State Chem. {\bf 163}, 498 (2002).

\bibitem{bobev}
S. Bobev, J. Merz, A. Lima, V. Fritsch, J.D. Thompson, J.L. Sarao,
M. Gillessen, and R. Dronskowski,
Inorg. Chem. {\bf 45}, 4047 (2006).

\bibitem{brock-bmp}
S.L. Brock, J.E. Greedan, and S.M. Kauzlarich,
J. Solid State Chem. {\bf 113}, 303 (1994).

\bibitem{xia-bms}
S.Q. Xia, C. Myers, and S. Bobev,
Eur. J. Inorg. Chem. {\bf 2008}, 4262 (2008).

\bibitem{sefat-s}
A.S. Sefat, M.A. McGuire, B.C. Sales, R.Y. Jin, J.Y. Howe, and D. Mandrus,
Phys. Rev. B {\bf 77}, 174503 (2008).

\bibitem{pinsard}
L. Pinsard-Gaudart, D. Berardan, J. Bobroff, and N. Dragoe,
Phys. Stat. Sol. Rapid Res. Lett. {\bf 2}, 185 (2008).

\bibitem{mcguire}
M.A. McGuire, A.C. Christianson, A.S. Sefat, B.C. Sales, M.D. Lumsden,
R.Y. Jin, E.A. Payzant, D. Mandrus, Y.B. Luan, V. Keppens, V. Varadarajan,
J.W. Brill, R.P. Hermann, M.T. Sougrati, F. Grandjean, and G.J. Long,
Phys. Rev. B {\bf 78}, 094517 (2008).

\bibitem{li}
L.J. Li, Y.K. Li, Z. Ren, Y.K. Luo, X. Lin, M. He, Q. Tao, Z.W. Zhu,
G.H. Cao, and Z.A. Xu,
Phys. Rev. B {\bf 78}, 132506 (2008).

\bibitem{li-ba}
L.J. Li, Y.K. Luo, Q.B. Wang, H. Chen, Z. Ren, Q. Tao, Y.K. Li, X. Lin,
M. He, Z.W. Zhu, G.H. Cao, and Z.A. Xu,
arXiv:0809.2009 (2008).

\bibitem{pbe}
J.P. Perdew, K. Burke, and M. Ernzerhof,
Phys. Rev. Lett. {\bf 77}, 3865 (1996).

\bibitem{singh-book}
D.J. Singh and L. Nordstrom, {\em Planewaves Pseudopotentials and the LAPW
Method, 2nd Ed.} (Springer, Berlin, 2006).

\bibitem{singh-lfao}
D.J. Singh and M.H. Du, Phys. Rev. Lett. {\bf 100}, 237003 (2008).

\bibitem{singh-bfa}
D.J. Singh, Phys. Rev. B {\bf 78}, 094511 (2008).

\bibitem{pwscf}
P. Giannozzi {\em et al.}, http://www.quantum-espresso.org.

\bibitem{wien2k}
P. Blaha, K. Schwarz, G. Madsen, D. Kvasnicka, and J. Luitz,
http://www.wien2k.at.

\bibitem{singh-lo}
D. Singh, Phys. Rev. B {\bf 43}, 6388 (1991).

\bibitem{boltztrap}
G.K.H. Madsen and D.J. Singh, Comput. Phys. Commun. {\bf 175}, 67 (2006).

\bibitem{hoffmann}
R. Hoffmann and C. Zheng, J. Phys. Chem. {\bf 89}, 4175 (1985).

\bibitem{zheng}
C. Zheng and R. Hoffmann,
J. Solid State Chem. {\bf 72}, 58 (1988).

\bibitem{gusteneau}
E. Gustenau, P. Herzig, and A. Neckel,
J. Alloys Compds. {\bf 262-263}, 516 (1997).

\bibitem{singh-fe}
D.J. Singh, Phys. Rev. B {\bf 45}, 2258 (1992).

\bibitem{pwa}
P.W. Anderson, Phys. Rev. {\bf 115}, 2 (1959).

\bibitem{matsuishi}
S. Matsuishi, Y. Inoue, T. Nomura, Y. Kamihara, M. Hirano, and H. Hosono,
arXiv:0811.1147 (2008).

\bibitem{terasaki}
I. Terasaki, Y. Sasago, and K. Uchinokura,
Phys. Rev. B {\bf 56}, R12685 (1997).

\bibitem{gga-note}
Our calculated thermopowers based on the GGA electronic structure
are similar except that the values at high temperatures and low
doping levels are larger
owing the the larger GGA gap, which leads to less minority carrier
conduction.

\end{references}
\end{document}